\def\be{\begin{equation}}
\def\ee{\end{equation}}
\def\ba{\begin{array}}
\def\ea{\end{array}}

\documentclass[prl,showpacs,twocolumn,amsmath]{revtex4}
\usepackage{amsfonts}
\usepackage{mathrsfs}
\usepackage{mathdots}
\usepackage{amssymb}
\usepackage{pifont}
\usepackage{epsfig,subfigure,dsfont,amsthm,amsbsy,mathrsfs,amscd}

\usepackage{rotating}

\usepackage{graphicx}
\usepackage{dcolumn}
\usepackage{bm}
\usepackage{amsmath}
\usepackage{multirow}
\usepackage{array}
\usepackage{epstopdf}
\usepackage{diagbox}
\usepackage{dcolumn}

\def\qed{\leavevmode\unskip\penalty9999 \hbox{}\nobreak\hfill
     \quad\hbox{\leavevmode  \hbox to.77778em{%
               \hfil\vrule   \vbox to.675em%
               {\hrule width.6em\vfil\hrule}\vrule\hfil}}
     \par\vskip3pt}

\begin{document}

\title{Quantumness near the Schwarzschild black hole based on W-state}
\author{Guang-Wei Mi$^{1}$}
\author{Xiaofen Huang $^{1}$}
\author{Shao-Ming Fei $^{2}$}
\author{Tinggui Zhang$^{1,*}$}
\affiliation{$^{1}$School of Mathematics and Statistics, Hainan Normal University, Haikou, 571158, China\\
$^{2}$School of Mathematical Sciences, Capital Normal University, Beijing, 100048, China\\
{$^{*}$ Email address: tinggui333@163.com}}


\begin{abstract}
We investigate certain quantumness in the vicinity of the Schwarzschild black hole by utilizing the W state. We explore the influence of the Hawking effect on the $l_{1}$-norm of quantum coherence, the first-order coherence (FOC), the concurrence-fill (CF) and the global concurrence (GC) in Schwarzschild black hole, for systems with one, two and three physically accessible modes. We conclude that the Hawking effect of the black hole not only disrupts but also enhance the quantum entanglement, while destroying the quantum coherence for systems with three physically accessible modes. For systems with one or two physically accessible modes, the Hawking effect exerts a positive influence on quantum coherence and quantum entanglement. Moreover, we study the influence of both the Hawking effect and environmental noise (AD channels) on $l_{1}$-norm of quantum coherence, FOC, CF and GC. It is demonstrated that
for systems with three physically accessible modes, the Hawking effect of the black hole disrupts quantum coherence but exerts a positive influence on quantum entanglement under the AD channels.
\end{abstract}

\pacs{03.67.Mn, 03.67.Hk}
\maketitle


\section{Introduction}
At the dawn of the 20th century, Einstein introduced the theory of general relativity, unveiling that gravity is a consequence of the curvature of spacetime. In 1916, the German physicist Karl Schwarzschild~\cite{Schwarzschild.1916} solved Einstein's field equations, leading to the discovery of the Schwarzschild radius surrounding a static, spherically symmetric celestial object. This radius signifies that if the physical radius of the celestial body is smaller than this critical value, even light cannot escape its gravitational pull. This groundbreaking finding marked the first rigorous mathematical demonstration of the existence of black holes. In 1974, Stephen Hawking put forward the theory of Hawking radiation~\cite{Hawking.1974}, which unveiled a profound connection between black holes and quantum mechanics and subsequently sparked the black hole information paradox~\cite{Hawking.1975,Hawking.1976,Bombelli.1986}. Stepping into the 21st century, humanity has achieved a monumental leap from theory to reality. The images of M87*~\cite{EHT.L1,EHT.L2,EHT.L3,EHT.L4,EHT.L5,EHT.L6} released in 2019 and the image of Sgr A*~\cite{EHT.L17} unveiled in 2022 have further validated the theory of general relativity.

Quantum coherence~\cite{Streltsov.2017,Memarzadeh.2017,Koyu.2018} secures high-efficiency, high-fidelity quantum information processing via superposition interference, whereas quantum entanglement~\cite{Bennett.1992,Bennett.1993,Horodecki.2009} facilitates long-haul communication and parallel computation through nonlocal correlations. Collectively, they form the cornerstone resources of quantum information technology. In 2014, Baumgratz \emph{et al.} quantified quantum coherence, thereby attracting widespread attention~\cite{Baumgratz.2014}. Besides, the first-order coherence (FOC) assumes critical importance in revealing quantum superpositions embedded within the local states of individual particles~\cite{Svozilik.2015,Ali.2021,Du.2021,D.D.Dong.2022}. Global entanglement and genuine multipartite entanglement (GME) are capable of unveiling more intricate multipartite correlations~\cite{Xie.2021}. The global concurrence (GC) emerged as a metric for quantifying global entanglement, providing a comprehensive measure of the total entanglement among all involved parties~\cite{Meyer.2002,Xie.2021}. Based on GC, Xie \emph{et al.} ~\cite{Xie.2021} proposed a measure, denoted as concurrence-fill (CF), for quantifying GME. Recently, Dong \emph{et al.}~\cite{D.D.Dong.2022} established a trade-off relationship between FOC and CF, thereby uncovering the intrinsic connection between quantum coherence and entanglement.

In recent years, the exploration of quantum information within non-inertial frames and curved spacetime has emerged as a burgeoning research frontier, driven by the quest to reconcile quantum mechanics and relativity~\cite{Schuller2005,Alsing2006,Wang2009,Wang2011,Esfahani2011,Xu2014,Kanno2016,Friis2016,Liu2018,Qiang2018,New2022,Eur2022,Li2022,Zhang.2023}. In~\cite{Ali.2024}, based on the GHZ state, Asad Ali \emph{et al.} delve into the quantum properties in the vicinity of a Schwarzschild black hole, examining diverse quantum resources and their interrelationships within the framework of curved spacetime. In~\cite{Wu.2023}, Wu \emph{et al.} investigate the genuine tripartite entanglement, one-tangle and two-tangle of W state of fermionic fields in the background of the Schwarzschild black hole. In~\cite{Kim.2022}, Kim \emph{et al.} explored the quantum entanglement and coherence properties of the W state for Dirac fields subjected to phase damping, phase flip and bit flip channels individually within non-inertial reference frames.

In the paper, we first discuss the influence of the Hawking effect on $l_{1}$-norm of quantum coherence, FOC, CF and GC in Schwarzschild black hole. Assume that Alice, Bob and Charlie initially share a W state at an asymptotically flat region. Let Bob and Charlie hover near the event horizon of the black hole, while Alice remain at the asymptotically flat region. We explore all the possible scenarios that encompass both physically accessible and inaccessible modes, that is, systems with three physically accessible modes, systems with two physically accessible modes, and systems with one physically accessible mode. For the first scenario, we conclude that the Hawking effect of the black hole not only disrupts the quantum entanglement but also enhance it, while destroying the quantum coherence. For the other two scenarios, the Hawking temperature $T$ exerts a positive influence on quantum coherence and quantum entanglement. We subsequently explore the influence of both the Hawking effect and environmental noise (AD channels) on $l_{1}$-norm of quantum coherence, FOC, CF and GC. For the first scenario, the Hawking effect of the black hole disrupts quantum coherence but exerts a positive influence on quantum entanglement under the AD channels. The conclusions for the other two scenarios are identical to those when no channels are present.

The rest of this paper is organized as follows. In Sec.\uppercase\expandafter{\romannumeral2}, we briefly reviewed the quantization of the Dirac field in a Schwarzschild black hole.
In Sec.\uppercase\expandafter{\romannumeral3}, we study the influence of the Hawking effect on $l_{1}$-norm of quantum coherence, FOC, CF and GC in Schwarzschild black hole.
In Sec. \uppercase\expandafter{\romannumeral4}, we investigate the influence of both the Hawking effect and environmental noise on $l_{1}$-norm of quantum coherence, FOC, CF and GC.
We conclude in Sec.\uppercase\expandafter{\romannumeral5}.

\section{QUANTIZATION OF DIRAC FIELD IN SCHWARZSCHILD BLACK HOLE}

The Dirac equation in a general spacetime can be expressed as~\cite{Dirac.1957},
\begin{eqnarray}
\begin{aligned}
\left[\gamma^{a}e^{\mu}_{a}(\partial_{\mu}+\Gamma_{\mu})\right]\Phi=0,
\end{aligned}
\end{eqnarray}
where $\gamma^{a}$ ($a=1,2,3$) are the Dirac matrices, the four-vector $e^{\mu}_{a}$ are the inverse of the tetrad $e^{a}_{\mu}$ and $\Gamma_{\mu}=\frac{1}{8}[\gamma^{a},\gamma^{b}]e^{\nu}_{a}e_{b\nu}$, with $\mu$ being the spin connection coefficients.
The metric describing the Schwarzschild spacetime is capable of being expressed as~\cite{metric.2010}
\begin{eqnarray}
\begin{aligned}
ds^{2}=&-\left(1-\frac{2M}{r}\right)dt^{2}+\left(1-\frac{2M}{r}\right)^{-1}dr^{2}\\[1mm]
&+r^{2}(d\theta^{2}+\sin^{2}\theta d\varphi^{2}),
\end{aligned}
\end{eqnarray}
where $M$ is the mass of the black hole. For simplicity, let $G=c=\hbar=k_{B}=1$. In particular, the Dirac equation in Schwarzschild spacetime can be written as~\cite{Xu.2014},
\begin{eqnarray}
\small
\label{SchDirac}
\begin{aligned}
&-\frac{\gamma^{1}}{\sqrt{1-\frac{2M}{r}}}\frac{\partial\Phi}{\partial t}+\gamma_{1}\sqrt{1-\frac{2M}{r}}\left[\frac{\partial}{\partial r}+\frac{1}{r}+\frac{M}{2r(r-2M)}\right]\Phi\\
&+\frac{\gamma^{2}}{r}\left(\frac{\partial}{\partial\theta}
+\frac{\cot\theta}{2}\right)\Phi
+\frac{\gamma^{3}}{r\sin\theta}\frac{\partial\Phi}{\partial\varphi}=0.
\end{aligned}
\end{eqnarray}

Eq.(\ref{SchDirac}) gives rise to positive (fermionic) frequency outgoing solutions in both regions inside and outside the event horizon~\cite{Jing.2004,Wang.2010,metric.2010},
\begin{eqnarray*}
\Phi^{+}_{k,out}=\xi e^{-i\omega\mu},~~
\Phi^{+}_{k,in}=\xi e^{i\omega\mu},
\end{eqnarray*}
where $k$ is the wave vector, $\xi$ represents the four-component Dirac spinor, $\omega$ indicates a monochromatic frequency and $\mu=t-r_{*}$ with the tortoise
coordinate $r_{*}=r+2M\ln\frac{r-2M}{2M}$.

A complete basis for the positive energy modes, i.e., the Kruskal modes, has been presented in~\cite{Damoar.1976}. By quantizing the Dirac field with respect to Schwarzschild and Kruskal modes, one derives the Bogoliubov transformation~\cite{Barnett.1997} connecting the creation and annihilation operators within Schwarzschild and Kruskal coordinates. Subsequently, the Kruskal vacuum and excited states in the Schwarzschild spacetime can be written as ~\cite{Wang2010,Wu2022,Phy2024},
\begin{eqnarray}
\begin{aligned}
\label{eqK}
&|0\rangle_{K}=\frac{1}{\sqrt{e^{-\frac{\omega}{T}}+1}}|0\rangle_{out}|0\rangle_{in}+\frac{1}{\sqrt{e^{\frac{\omega}{T}}+1}}|1\rangle_{out}|1\rangle_{in},\\
&|1\rangle_{K}=|1\rangle_{out}|0\rangle_{in},
\end{aligned}
\end{eqnarray}
where $T=\frac{1}{8\pi M}$ is the Hawking temperature, $|n\rangle_{out}$ and $|n\rangle_{in}$ correspond to the fermionic modes outside the event horizon and the antifermionic modes inside the event horizon, respectively.

\section{The Hawking effect on $l_{1}$-norm of quantum coherence, FOC, CF and GC in Schwarzschild black hole}

In this section, we investigate certain quantumness in the vicinity of a Schwarzschild black hole by utilizing the W state. Initially, we suppose that Alice, Bob and Charlie share a W state at an asymptotically flat region, which can be written as
\begin{eqnarray}
\begin{aligned}
\label{Wstate}
|W\rangle_{ABC}=\frac{1}{\sqrt{3}}(|001\rangle+|010\rangle+|100\rangle).
\end{aligned}
\end{eqnarray}
Consider the scenario in which Alice remains at the asymptotically flat region, while Bob and Charlie hover in close proximity to the event horizon of black hole. Therefore, from Eq.(\ref{eqK}), Eq.(\ref{Wstate}) can be rewritten as
\begin{eqnarray}
\begin{aligned}
\label{W-Hstate}
|W\rangle_{ABbCc}&=\frac{1}{\sqrt{3}}\left[\alpha|00010\rangle+\beta|01110\rangle+\alpha|01000\rangle\right.\\
&\left.+\beta|01011\rangle+\alpha^{2}|10000\rangle+\beta^{2}|11111\rangle\right.\\
&\left.+\alpha\beta(|10011\rangle+|11100\rangle)\right],
\end{aligned}
\end{eqnarray}
where $\alpha=\frac{1}{\sqrt{e^{-\frac{\omega}{T}}+1}}$ and $\beta=\frac{1}{\sqrt{e^{\frac{\omega}{T}}+1}}$, $B$ and $C$ represent the fermionic particles located outside the event horizon of a black hole, respectively, while $b$ and $c$ denote the antifermionic particles situated within the event horizon, respectively.
Generally, the fermionic particles residing outside the event horizon of a black hole are treated as physically accessible modes, whereas the antifermionic particles confined within the horizon are classified as physically inaccessible modes in the theoretical formalism. Consequently, with respect to (\ref{W-Hstate}), the fermionic modes $A$, $B$ and $C$ are classified as physically accessible modes, while the antifermionic modes $b$ and $c$ are rigorously identified as physically inaccessible modes. We explore all possible scenarios that encompass both physically accessible and inaccessible modes by examining the following scenarios:

$\textbf{Case(1): Systems with three physically access-}$ $\textbf{ible modes.}$ We first consider the scenario with three physically accessible modes, i.e., the fermionic particles $A$, $B$ and $C$. By tracing over the physically inaccessible modes $b$ and $c$, from (\ref{W-Hstate}) we obtain the state
\begin{eqnarray}
\begin{aligned}
\label{state-ABC}
\rho_{ABC}&=\frac{1}{3}\left[\alpha^{2}(|001\rangle\langle001|+|001\rangle\langle010|+|010\rangle\langle001|\right.\\
&\left.+|010\rangle\langle010|)+\alpha^{3}(|001\rangle\langle100|+|100\rangle\langle001|\right.\\
&\left.+|010\rangle\langle100|+|100\rangle\langle010|)+2\beta^{2}|011\rangle\langle011|\right.\\
&\left.+\alpha\beta^{2}(|011\rangle\langle110|+|110\rangle\langle011|+|011\rangle\langle101|\right.\\
&\left.+|101\rangle\langle011|)+\alpha^{2}\beta^{2}(|101\rangle\langle101|+|110\rangle\langle110|)\right.\\
&\left.+\alpha^{4}|100\rangle\langle100|+\beta^{4}|111\rangle\langle111|\right].
\end{aligned}
\end{eqnarray}
The matrix representation of the state $\rho_{ABC}$ is given by
\begin{eqnarray}
\label{rhoABC}
\rho_{ABC}=\frac{1}{3}
\left (
\begin{array}{cccccccc}
0       &0                &0                  &0                 &0                &0                     &0                       &0    \\
0       &\alpha^{2}       &\alpha^{2}         &0                 &\alpha^{3}       &0                     &0                       &0    \\
0       &\alpha^{2}       &\alpha^{2}         &0                 &\alpha^{3}       &0                     &0                       &0    \\
0       &0                &0                  &2\beta^{2}        &0                &\alpha\beta^{2}       &\alpha\beta^{2}         &0    \\
0       &\alpha^{3}       &\alpha^{3}         &0                 &\alpha^{4}       &0                     &0                       &0    \\
0       &0                &0                  &\alpha\beta^{2}   &0                &\alpha^{2}\beta^{2}   &0                       &0    \\
0       &0                &0                  &\alpha\beta^{2}   &0                &0                     &\alpha^{2}\beta^{2}     &0    \\
0       &0                &0                  &0                 &0                &0                     &0                       &\beta^{4}
\end{array}
\right ).
\end{eqnarray}

Subsequently, we investigate the influence of the Hawking effect on $l_{1}$-norm of quantum coherence, first-order coherence, concurrence fill and global concurrence in Schwarzschild black hole. The $l_{1}$-norm of quantum coherence of a state $\rho$ is defined by \cite{Baumgratz.2014}
\begin{eqnarray}
\begin{aligned}
\label{l1-norm}
C_{l_{1}}(\rho)=\sum_{i\neq j}|\rho_{ij}|.
\end{aligned}
\end{eqnarray}
Therefore, from Eq.(\ref{rhoABC}) and Eq.(\ref{l1-norm}), we have
\begin{eqnarray}
\begin{aligned}
\label{l1ABC}
C_{l_{1}}(\rho_{ABC})=\frac{2}{3}(\alpha^{2}+2\alpha).
\end{aligned}
\end{eqnarray}

Another measure FOC of quantum coherence for a tripartite states $\rho_{xyz}$ is given by~\cite{D.D.Dong.2022}
\begin{eqnarray}
\begin{aligned}
\label{FOC}
D(\rho_{xyz})=\sqrt{\frac{D^{2}(\rho_{x})+D^{2}(\rho_{y})+D^{2}(\rho_{z})}{3}},
\end{aligned}
\end{eqnarray}
where $D(\rho_{i})=\sqrt{2tr(\rho_{i}^{2})-1}$ with $\rho_{i}$ ($i\in\{x,y,z\}$) being the reduced state associated with the $i$th subsystem. It is verified that $0\leq D(\rho_{xyz})\leq1$.
From Eq.(\ref{rhoABC}) and Eq.(\ref{FOC}), we have
\begin{eqnarray}
\begin{aligned}
\label{ABC-FOC}
D(\rho_{ABC})=\frac{1}{3}\sqrt{\frac{16(\alpha^{4}+\beta^{2}+\beta^{4})-13}{3}},
\end{aligned}
\end{eqnarray}
see the details in Appendix A.

For tripartite pure states $\rho_{xyz}$, the CF is denoted as~\cite{Xie.2021}
\begin{eqnarray}
\begin{aligned}
\label{CF}
F(\rho_{xyz})&=\left\{\frac{16}{3}Q(\rho_{xyz})\times\left[Q(\rho_{xyz})
-C^{2}_{x(yz)}(\rho_{xyz})\right] \right.\\
&\left.\times\left[Q(\rho_{xyz})-C^{2}_{y(zx)}(\rho_{xyz})\right] \right.\\
&\left.\times\left[Q(\rho_{xyz})-C^{2}_{z(xy)}(\rho_{xyz})\right] \right\}^{\frac{1}{4}},
\end{aligned}
\end{eqnarray}
where
\begin{eqnarray}
\small
\begin{aligned}
\label{GC}
Q(\rho_{xyz})=\frac{1}{2}\left[C^{2}_{x(yz)}(\rho_{xyz})
+C^{2}_{y(zx)}(\rho_{xyz})+C^{2}_{z(xy)}(\rho_{xyz})\right]
\end{aligned}
\end{eqnarray}
is named as GC~\cite{Meyer.2002,Xie.2021}, $C_{i(jk)}=2\sqrt{det(\rho_{i})}$
($i\neq j\neq k\in\{x,y,z\}$) and $0\leq C_{i(jk)}\leq1$. The prefactor $\frac{16}{3}$ ensures the normalization $0\leq F(\rho_{xyz})\leq1$.
From Eq.(\ref{rhoABC}), Eq.(\ref{CF}) and Eq.(\ref{GC}), we deduce that (see Appendix A)
\begin{eqnarray}
\begin{aligned}
\label{ABC-CF}
F(\rho_{ABC})=\frac{4}{3}\left\{\frac{1}{3}Q(\rho_{ABC})\times\left[Q(\rho_{ABC})-\frac{8}{9}\right]\right\}^{\frac{1}{4}},
\end{aligned}
\end{eqnarray}
where
\begin{eqnarray}
\begin{aligned}
\label{ABC-GC}
Q(\rho_{ABC})=\frac{4}{9}\left[1+2\alpha^{2}(1+2\beta^{2})\right].
\end{aligned}
\end{eqnarray}
A trade-off relation between FOC and CF has been established in~\cite{D.D.Dong.2022},
\begin{eqnarray}
\begin{aligned}
\label{FOC-CF}
D^{2}(\rho_{xyz})+F(\rho_{xyz})\leq1.
\end{aligned}
\end{eqnarray}

\begin{figure}[h]
\centering
\includegraphics[scale=0.62]{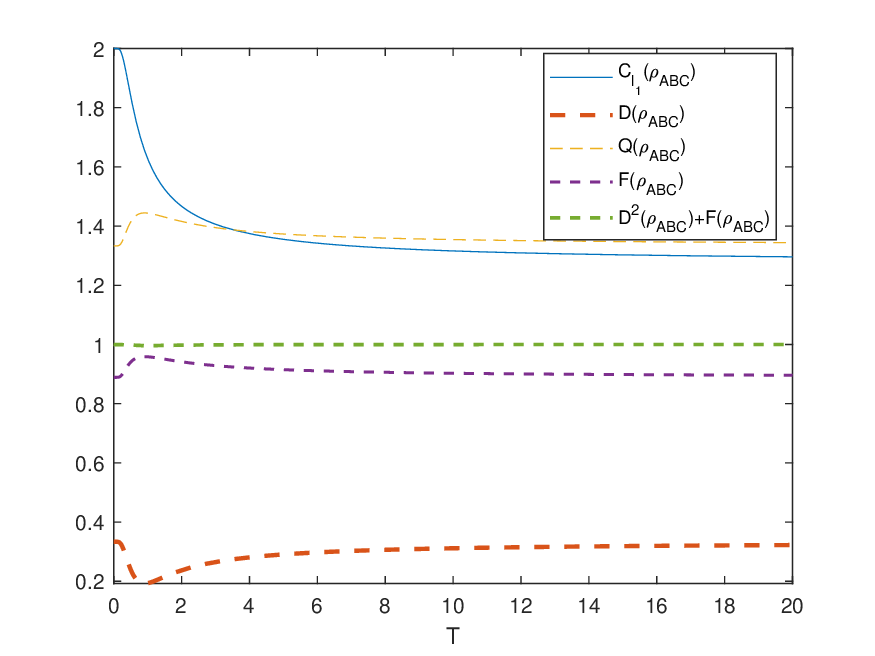}
\caption{The $l_{1}$-norm of quantum coherence $[C_{l_{1}}(\rho_{ABC})]$, FOC $[D(\rho_{ABC})]$, GC $[Q(\rho_{ABC})]$, CF $[F(\rho_{ABC})]$, and $D^{2}(\rho_{ABC})+F(\rho_{ABC})$ as functions of the Hawking temperature $T$ with $\omega=1$.}
\label{FigABC}
\end{figure}

From Eq.(\ref{l1ABC}), Eq.(\ref{ABC-FOC}), Eq.(\ref{ABC-CF}) and Eq.(\ref{ABC-GC}), we observe that the aforementioned quantities are correlated to the Hawking temperature $T$. This implies that Hawking radiation would influence the coherence and entanglement in the Schwarzschild black hole. In Fig.\ref{FigABC}, we depict the $l_{1}$-norm of quantum coherence, FOC, GC, CF and $D^{2}(\rho_{ABC})+F(\rho_{ABC})$ as functions of the Hawking temperature $T$ with $\omega=1$. It is seen that $C_{l_{1}}(\rho_{ABC})$ decreases monotonically as the Hawking temperature $T$ increases. $C_{l_{1}}(\rho_{ABC})$ gradually diminishes from its maximum value $2$ and eventually stabilizes. $Q(\rho_{ABC})$, $F(\rho_{ABC})$ and $D(\rho_{ABC})$ exhibit non-monotonic variations with respect to $T$. Specifically, $Q(\rho_{ABC})$ and $F(\rho_{ABC})$ initially increase gradually to maximum before decreasing progressively, while $D(\rho_{ABC})$ first decreases gradually to a minimum and then increases steadily. Ultimately, $Q(\rho_{ABC})$, $F(\rho_{ABC})$ and $D(\rho_{ABC})$ stabilize at non-zero values.
Notably, a nearly perfect trade-off relationship has been essentially achieved between $F(\rho_{ABC})$ and $D(\rho_{ABC})$, that is, $D^{2}(\rho_{xyz})+F(\rho_{xyz})\approx1$.

Therefore, from Fig.\ref{FigABC} we conclude that the Hawking effect of the black hole not only disrupts but also enhance the quantum entanglement in a completely physically accessible scenario, while destroying the quantum coherence. Moreover, even the Hawking temperature tends to infinity, the quantum entanglement and quantum coherence cannot be entirely eradicated.
This result diverges from the conclusion in~\cite{Ali.2024}. In~\cite{Ali.2024}, the authors claimed that Hawking temperatures degrade entanglement based on GHZ states for a completely accessible scenario. Interestingly, our research reveals that the result based on the W state and the GHZ state are distinct.

$\textbf{Case(2): Systems with one physically accessib-}$ $\textbf{le mode and two inaccessible modes.}$ Consider the scenario with fermionic particle $A$, and antifermionic particles $b$ and $c$. By tracing out the physically accessible modes $B$ and $C$, we have from (\ref{W-Hstate})
\begin{eqnarray}
\label{rhoAbc}
\rho_{Abc}=\frac{1}{3}
\left (
\begin{array}{cccccccc}
2\alpha^{2}       &0                &0                  &0                 &0                &\alpha^{2}\beta       &\alpha^{2}\beta         &0            \\
0                 &\beta^{2}        &\beta^{2}          &0                 &0                &0                     &0                       &\beta^{3}    \\
0                 &\beta^{2}        &\beta^{2}          &0                 &0                &0                     &0                       &\beta^{3}    \\
0                 &0                &0                  &0                 &0                &0                     &0                       &0             \\
0                 &0                &0                  &0                 &\alpha^{4}       &0                     &0                       &0             \\
\alpha^{2}\beta   &0                &0                  &0                 &0                &\alpha^{2}\beta^{2}   &0                       &0            \\
\alpha^{2}\beta   &0                &0                  &0                 &0                &0                     &\alpha^{2}\beta^{2}     &0           \\
0                 &\beta^{3}        &\beta^{3}          &0                 &0                &0                     &0                       &\beta^{4}
\end{array}
\right).
\end{eqnarray}
From Eq.(\ref{l1-norm}), Eq.(\ref{FOC}), Eq.(\ref{CF}), Eq.(\ref{GC}) and Eq.(\ref{rhoAbc}), we deduce that (see Appendix A)
\begin{eqnarray}
\begin{aligned}
\label{Abc-QC}
C_{l_{1}}(\rho_{Abc})=\frac{2}{3}(\beta^{2}+2\beta),
\end{aligned}
\end{eqnarray}
\begin{eqnarray}
\begin{aligned}
\label{Abc-FOC}
D(\rho_{Abc})=\frac{1}{3}\sqrt{\frac{16(\alpha^{4}+\alpha^{2}+\beta^{4})-13}{3}},
\end{aligned}
\end{eqnarray}
\begin{eqnarray}
\begin{aligned}
\label{Abc-CF}
F(\rho_{Abc})=\frac{4}{3}\left\{\frac{1}{3}Q(\rho_{Abc})\times\left[Q(\rho_{Abc})-\frac{8}{9}\right]\right\}^{\frac{1}{4}},
\end{aligned}
\end{eqnarray}
where
\begin{eqnarray}
\begin{aligned}
\label{Abc-GC}
Q(\rho_{Abc})=\frac{4}{9}\left[1+2\beta^{2}(1+2\alpha^{2})\right].
\end{aligned}
\end{eqnarray}

\begin{figure}[ht]
\centering
\includegraphics[scale=0.62]{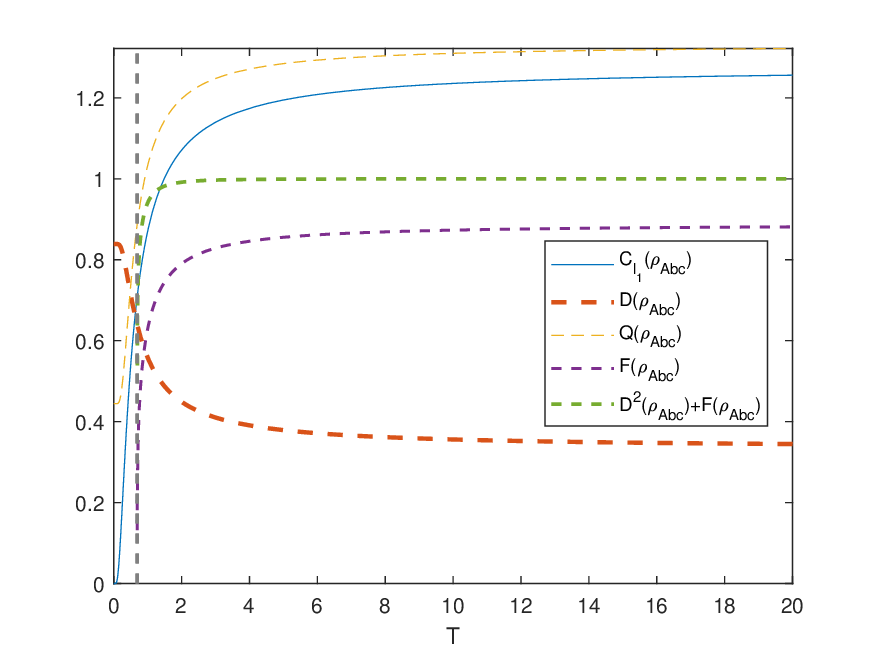}
\caption{The $l_{1}$-norm of quantum coherence $[C_{l_{1}}(\rho_{Abc})]$, FOC $[D(\rho_{Abc})]$, GC $[Q(\rho_{Abc})]$, CF $[F(\rho_{Abc})]$, and $D^{2}(\rho_{Abc})+F(\rho_{Abc})$ as functions of the Hawking temperature $T$ with $\omega=1$.}
\label{FigAbc}
\end{figure}

Fig.\ref{FigAbc} shows $C_{l_{1}}(\rho_{Abc})$, $D(\rho_{Abc})$, $Q(\rho_{Abc})$, $F(\rho_{Abc})$ and $D^{2}(\rho_{Abc})+F(\rho_{Abc})$ as functions of the Hawking temperature $T$ with $\omega=1$. Evidently, we observe that the $C_{l_{1}}(\rho_{Abc})$, $Q(\rho_{Abc})$ and $F(\rho_{Abc})$ increases monotonically with the rising Hawking temperature $T$, while $D(\rho_{Abc})$ decreases monotonically as Hawking temperature $T$ increases. $C_{l_{1}}(\rho_{Abc})$, $Q(\rho_{Abc})$ and $F(\rho_{Abc})$ gradually increase from their respective minima and asymptotically approach stable values, whereas $D(\rho_{Abc})$ progressively decreases from its maximum and stabilizes then.
Additionally, $C_{l_{1}}(\rho_{Abc})_{min}=0$, $Q(\rho_{Abc})_{min}=\frac{4}{9}$ and $D(\rho_{Abc})_{max}=\sqrt{\frac{19}{27}}$. Besides, from Eq.(\ref{Abc-CF}), $F(\rho_{Abc})$ is well-defined only when $Q(\rho_{Abc})\geq\frac{8}{9}$ and $F(\rho_{Abc})=0$ when $Q(\rho_{Abc})=\frac{8}{9}$. $D^{2}(\rho_{Abc})+F(\rho_{Abc})$ monotonically increases and converges to $1$ as $T$ increases. In other words, a near-optimal trade-off relationship is attained between $F(\rho_{ABC})$ and $D(\rho_{ABC})$ with the rising Hawking temperature $T$.
From Fig.\ref{FigAbc}, we come to the conclusion that the Hawking effect exerts a positive influence on quantum coherence and entanglement.

$\textbf{Case(3): Systems with two physically accessib-}$ $\textbf{le modes and one inaccessible mode.}$ We now consider the scenario with fermionic particles $A$ and $B$ and antifermionic particle $c$. From Eq.(\ref{W-Hstate}) one acquires by tracing over the physically inaccessible mode $b$ and physically accessible mode $C$,
\begin{eqnarray}
\label{rhoABc}
\rho_{ABc}=\frac{1}{3}
\left (
\begin{array}{cccccccc}
\alpha^{2}        &0                &0                  &\alpha\beta       &0                &\alpha^{2}\beta       &0                       &0            \\
0                 &0                &0                  &0                 &0                &0                     &0                       &0            \\
0                 &0                &1                  &0                 &\alpha^{3}       &0                     &0                       &\beta^{3}    \\
\alpha\beta       &0                &0                  &\beta^{2}         &0                &\alpha\beta^{2}       &0                       &0             \\
0                 &0                &\alpha^{3}         &0                 &\alpha^{4}       &0                     &0                       &0             \\
\alpha^{2}\beta   &0                &0                  &\alpha\beta^{2}   &0                &\alpha^{2}\beta^{2}   &0                       &0            \\
0                 &0                &0                  &0                 &0                &0                     &\alpha^{2}\beta^{2}     &0           \\
0                 &0                &\beta^{3}          &0                 &0                &0                     &0                       &\beta^{4}
\end{array}
\right ).
\end{eqnarray}
From Eq.(\ref{l1-norm}), Eq.(\ref{FOC}), Eq.(\ref{CF}), Eq.(\ref{GC}) and Eq.(\ref{rhoABc}), we have (see Appendix A)
\begin{eqnarray}
\begin{aligned}
\label{ABc-QC}
C_{l_{1}}(\rho_{ABc})=\frac{2}{3}(\alpha+\beta+\alpha\beta),
\end{aligned}
\end{eqnarray}
\begin{eqnarray}
\begin{aligned}
\label{ABc-FOC}
D(\rho_{ABc})=\frac{1}{3}\sqrt{\frac{16(\alpha^{4}+\beta^{4})-5}{3}},
\end{aligned}
\end{eqnarray}
\begin{eqnarray}
\begin{aligned}
\label{ABc-CF}
F(\rho_{ABc})=\frac{4}{3}\alpha\beta\left[\frac{64}{27}Q(\rho_{ABc})\right]^{\frac{1}{4}},
\end{aligned}
\end{eqnarray}
where
\begin{eqnarray}
\begin{aligned}
\label{ABc-GC}
Q(\rho_{ABc})=\frac{8}{9}\left(1+2\alpha^{2}\beta^{2}\right).
\end{aligned}
\end{eqnarray}

\begin{figure}[h]
\centering
\includegraphics[scale=0.62]{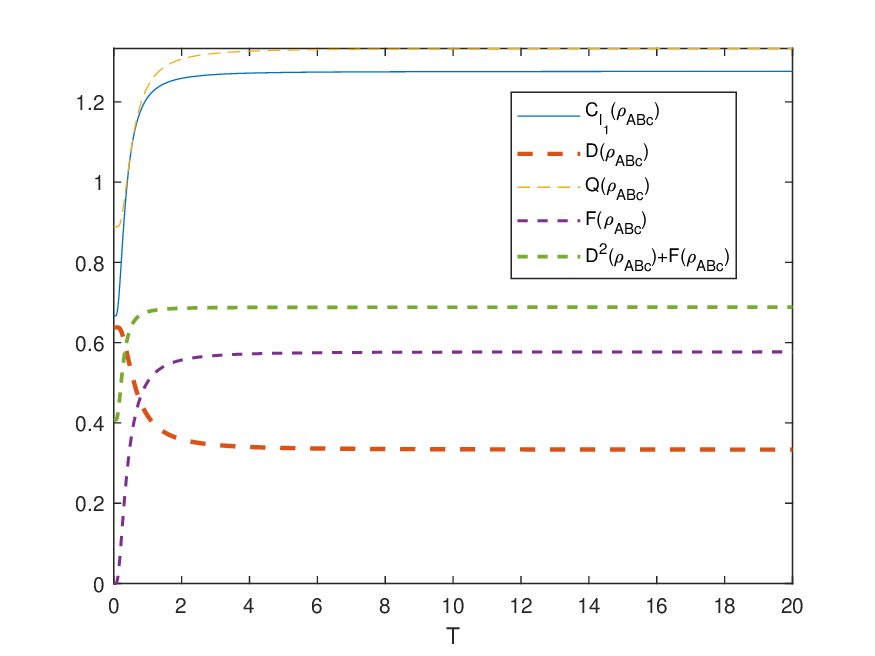}
\caption{The $l_{1}$-norm of quantum coherence $[C_{l_{1}}(\rho_{ABc})]$, FOC $[D(\rho_{ABc})]$, GC $[Q(\rho_{ABc})]$, CF $[F(\rho_{ABc})]$, and $D^{2}(\rho_{ABc})+F(\rho_{ABc})$ as functions of the Hawking temperature $T$ with $\omega=1$.}
\label{FigABc}
\end{figure}

In Fig.\ref{FigABc}, we explore the $C_{l_{1}}(\rho_{ABc})$, $D(\rho_{ABc})$, $Q(\rho_{ABc})$, $F(\rho_{ABc})$, and $D^{2}(\rho_{ABc})+F(\rho_{ABc})$ as functions of the Hawking temperature $T$ with $\omega=1$. It observed that $C_{l_{1}}(\rho_{ABc})$, $Q(\rho_{ABc})$ and $F(\rho_{ABc})$ exhibit a monotonic growth trend with respect to increasing Hawking temperature $T$. Conversely, $D(\rho_{ABc})$ demonstrates a strictly decreasing dependency on $T$.
$C_{l_{1}}(\rho_{ABc})$, $Q(\rho_{ABc})$ and $F(\rho_{ABc})$ monotonically increase from their respective minimum values of $\frac{2}{3}$, $\frac{8}{9}$ and $0$, while $D(\rho_{ABc})$ monotonically decreases from its maximum value of $\sqrt{\frac{11}{27}}$. Eventually, all of them stabilize. In addition, as the Hawking temperature $T$ rises, the trade-off relationship between $D(\rho_{ABc})$ and $F(\rho_{ABc})$ intensifies in a monotonic manner. The minimum value of $D^{2}(\rho_{ABc})+F(\rho_{ABc})$ is $\frac{11}{27}$. As the Hawking temperature $T$ increases incrementally, this trade-off relationship between $D(\rho_{ABc})$ and $F(\rho_{ABc})$ remains stable. Thus, in this case the Hawking effect of the black hole enhances the quantum coherence and entanglement.

\section{The Hawking effect on $l_{1}$-norm of quantum coherence, FOC, CF and GC under AD channels}

The amplitude damping (AD) channel $\varepsilon$ characterizes the phenomenon of spontaneous quantum state decay driven by energy dissipation when a quantum system interacts with environments~\cite{Nielsen.2000},
\begin{eqnarray}
\begin{aligned}
\varepsilon(\rho)=E_{0}\rho E_{0}^{\dag}+E_{1}\rho E_{1}^{\dag},
\end{aligned}
\end{eqnarray}
where the Kraus operators
\begin{eqnarray}\label{E0E1}
E_{0}=
\left (
\begin{array}{cc}
1                 &0                \\
0                 &\sqrt{1-\gamma}
\end{array}
\right ),
~~~E_{1}=
\left (
\begin{array}{cc}
0                 &\sqrt{\gamma}   \\
0                 &0
\end{array}
\right )
\end{eqnarray}
with $0\leq\gamma\leq1$.

For simplicity, we assume that the AD channels associated with systems Alice, Bob and Charlie are identical, that is, $\gamma_{A}=\gamma_{B}=\gamma_{C}$.
The complete temporal evolution of the three-qubit system subjected to AD channels can be formally expressed as~\cite{Salles.2000,Liu.2024},
\begin{eqnarray}
\begin{aligned}
\label{ADchannel}
\rho'=\sum^{1}_{i,j,k=0}E_{i}\otimes E_{j}\otimes E_{k}\rho E_{i}^{\dag}\otimes E_{j}^{\dag}\otimes E_{k}^{\dag}.
\end{aligned}
\end{eqnarray}
The states of a single subsystem after AD channels are given in Appendix B.

Firstly, we consider the physically accessible density matrix $\rho'_{ABC}$. From Eq.(\ref{rhoABC}) and  Eq.(\ref{ADchannel}), we have
\begin{eqnarray}
\label{rho'ABC}
\rho'_{ABC}=\frac{1}{3}
\left (
\begin{array}{cccccccc}
a_{11}       &0                &0              &0                 &0                &0                     &0                       &0    \\
0            &a_{22}           &a_{23}         &0                 &a_{25}           &0                     &0                       &0    \\
0            &a_{32}           &a_{33}         &0                 &a_{35}           &0                     &0                       &0    \\
0            &0                &0              &a_{44}            &0                &a_{46}                &a_{47}                  &0    \\
0            &a_{52}           &a_{53}         &0                 &a_{55}           &0                     &0                       &0    \\
0            &0                &0              &a_{64}            &0                &a_{66}                &0                       &0    \\
0            &0                &0              &a_{74}            &0                &0                     &a_{77}                  &0    \\
0            &0                &0              &0                 &0                &0                     &0                       &a_{88}
\end{array}
\right )
\end{eqnarray}
with
\begin{eqnarray}
\begin{aligned}
&a_{11}=\gamma(\alpha^{2}+\beta^{2}\gamma)^{2}+2\alpha^{2}\gamma+2\beta^{2}\gamma^{2},\\
&a_{22}=a_{33}=\alpha^{2}(1-\gamma)+\gamma(1-\gamma)\beta^{2}(\alpha^{2}+\beta^{2}\gamma+2),\\
&a_{44}=\beta^{2}(1-\gamma)^{2}(\beta^{2}\gamma+2),\\
&a_{55}=(1-\gamma)(\alpha^{2}+\beta^{2}\gamma)^{2},\\
&a_{66}=a_{77}=(1-\gamma)^{2}\beta^{2}(\alpha^{2}+\beta^{2}\gamma),\\
&a_{88}=(1-\gamma)^{3}\beta^{4},\\
&a_{23}=a_{32}=(1-\gamma)\alpha^{2},\\
&a_{25}=a_{52}=a_{35}=a_{53}=(1-\gamma)\alpha(\alpha^{2}+\beta^{2}\gamma),\\
&a_{46}=a_{64}=a_{47}=a_{74}=(1-\gamma)^{2}\alpha\beta^{2},
\end{aligned}
\end{eqnarray}
where $\alpha={1}/{\sqrt{e^{-\frac{\omega}{T}}+1}}$ and $\beta={1}/{\sqrt{e^{\frac{\omega}{T}}+1}}$.

\begin{figure}[htbp]
    \centering
    \begin{minipage}[b]{0.4\textwidth} 
        \includegraphics[width=\linewidth]{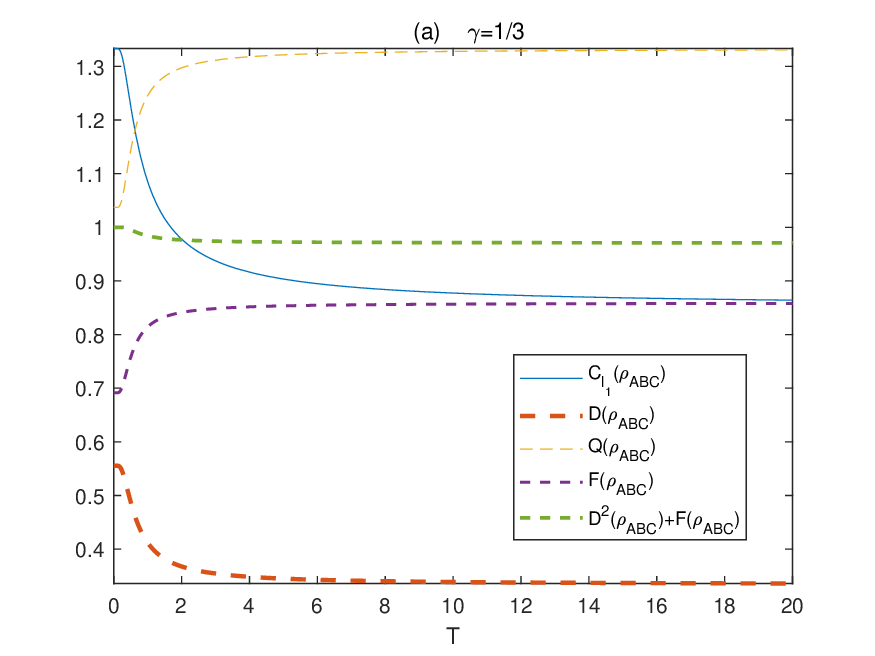} 

    \end{minipage}
    \hfill
    \begin{minipage}[b]{0.4\textwidth}
        \includegraphics[width=\linewidth]{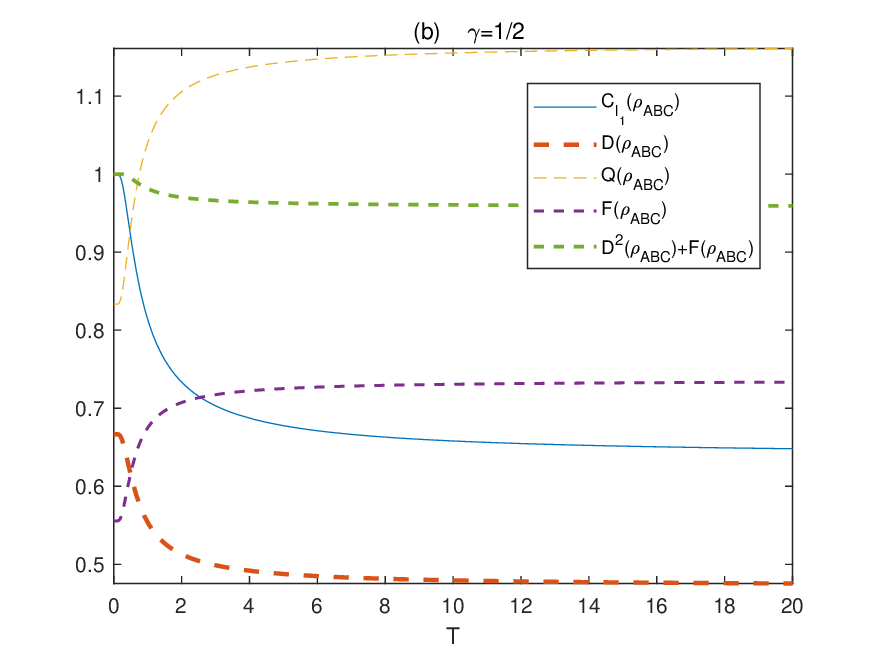}
    \end{minipage}
    \hfill
    \begin{minipage}[b]{0.4\textwidth}
        \includegraphics[width=\linewidth]{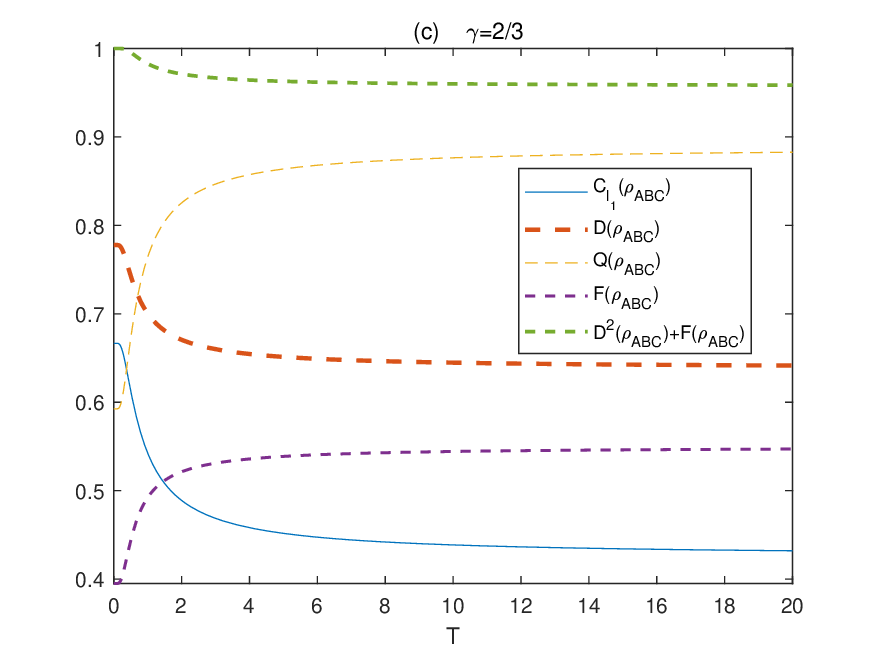}
    \end{minipage}
    \caption{The $l_{1}$-norm of quantum coherence $[C_{l_{1}}(\rho'_{ABC})]$, FOC $[D(\rho'_{ABC})]$, GC $[Q(\rho'_{ABC})]$, CF $[F(\rho'_{ABC})]$, and $D^{2}(\rho'_{ABC})+F(\rho'_{ABC})$ as functions of the Hawking temperature $T$ with $\omega=1$ for different $\gamma$: $\gamma=1/3$, $\gamma=1/2$ and $\gamma=2/3$.}
    \label{ADFigABC}
\end{figure}

In Fig.\ref{ADFigABC} (a-c), we illustrate $C_{l_{1}}(\rho'_{ABC})$, $D(\rho'_{ABC})$, $Q(\rho'_{ABC})$, $F(\rho'_{ABC})$ and $D^{2}(\rho'_{ABC})+F(\rho'_{ABC})$ as functions of the Hawking temperature $T$ with $\omega=1$ for different $\gamma$. For $\gamma=\frac{1}{3}$, from Fig.\ref{ADFigABC}(a), we see that $C_{l_{1}}(\rho'_{ABC})$ and $D(\rho'_{ABC})$ decrease monotonically as the Hawking temperature $T$ increases, while those of $Q(\rho'_{ABC})$ and $F(\rho'_{ABC})$ increase monotonically with the increasing Hawking temperature $T$. Interestingly, when $T\rightarrow0$, a perfect trade-off relationship is attained between $D(\rho'_{ABC})$ and $F(\rho'_{ABC})$, i.e., $D^{2}(\rho'_{ABC})+F(\rho'_{ABC})=1$. This trade-off relationship then gradually decreases as the Hawking temperature $T$ rises, eventually approaching a stable value. Besides, when $T\rightarrow0$, $C_{l_{1}}(\rho'_{ABC})=\frac{4}{3}$, $D(\rho'_{ABC})=\frac{5}{9}$, $Q(\rho'_{ABC})=\frac{28}{27}$ and $F(\rho'_{ABC})=\frac{56}{81}$. They reached their respective extreme values.

For $\gamma=\frac{1}{2}$ and $\gamma=\frac{2}{3}$, from Fig.\ref{ADFigABC}(b) and (c) we observe that the trends are identical to those depicted in Fig.\ref{ADFigABC}(a), but the actual values differ. When the Hawking temperature $T$ remains constant, $C_{l_{1}}(\rho'_{ABC})$, $Q(\rho'_{ABC})$ and $F(\rho'_{ABC})$ decrease as $\gamma$ increases, whereas $D(\rho'_{ABC})$ increases as $\gamma$ increases. Therefore, we may conclude that the Hawking effect of the black hole disrupts quantum coherence but exerts a positive influence on quantum entanglement under the AD channels in the completely physically accessible scenario.

Next, we consider the density matrix $\rho'_{Abc}$ under the AD channels. By Eq.(\ref{rhoAbc}) and  Eq.(\ref{ADchannel}), $\rho'_{Abc}$ can be expressed as
\begin{eqnarray}
\label{rho'ABC}
\rho'_{Abc}=\frac{1}{3}
\left (
\begin{array}{cccccccc}
b_{11}       &0                &0              &0                 &0                &b_{16}                &b_{17}                  &0        \\
0            &b_{22}           &b_{23}         &0                 &0                &0                     &0                       &b_{28}    \\
0            &b_{32}           &b_{33}         &0                 &0                &0                     &0                       &b_{38}    \\
0            &0                &0              &b_{44}            &0                &0                     &0                       &0         \\
0            &0                &0              &0                 &b_{55}           &0                     &0                       &0         \\
b_{61}       &0                &0              &0                 &0                &b_{66}                &0                       &0         \\
b_{71}       &0                &0              &0                 &0                &0                     &b_{77}                  &0          \\
0            &b_{82}           &b_{83}         &0                 &0                &0                     &0                       &b_{88}
\end{array}
\right )
\end{eqnarray}
with
\begin{eqnarray}
\begin{aligned}
&b_{11}=\gamma(\alpha^{2}+\beta^{2}\gamma)^{2}+2\alpha^{2}+2\beta^{2}\gamma,\\
&b_{22}=b_{33}=\beta^{2}(1-\gamma)(\alpha^{2}\gamma+\beta^{2}\gamma^{2}+1),\\
&b_{44}=\gamma(1-\gamma)^{2}\beta^{4},\\
&b_{55}=(1-\gamma)(\alpha^{2}+\beta^{2}\gamma)^{2},\\
&b_{66}=b_{77}=(1-\gamma)^{2}\beta^{2}(\alpha^{2}+\beta^{2}\gamma),\\
&b_{88}=(1-\gamma)^{3}\beta^{4},\\
&b_{23}=b_{32}=(1-\gamma)\beta^{2},\\
&b_{16}=b_{61}=b_{17}=b_{71}=(1-\gamma)\beta(\alpha^{2}+\beta^{2}\gamma),\\
&b_{28}=b_{82}=b_{38}=b_{83}=(1-\gamma)^{2}\beta^{3}.
\end{aligned}
\end{eqnarray}

\begin{figure}[htbp]
    \centering
    \begin{minipage}[b]{0.4\textwidth} 
        \includegraphics[width=\linewidth]{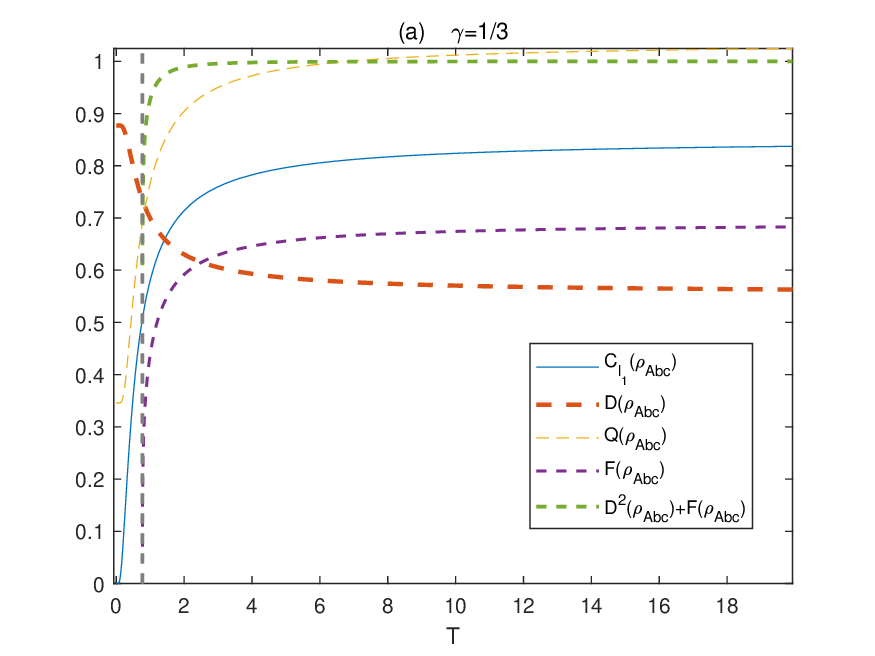} 

    \end{minipage}
    \hfill
    \begin{minipage}[b]{0.4\textwidth}
        \includegraphics[width=\linewidth]{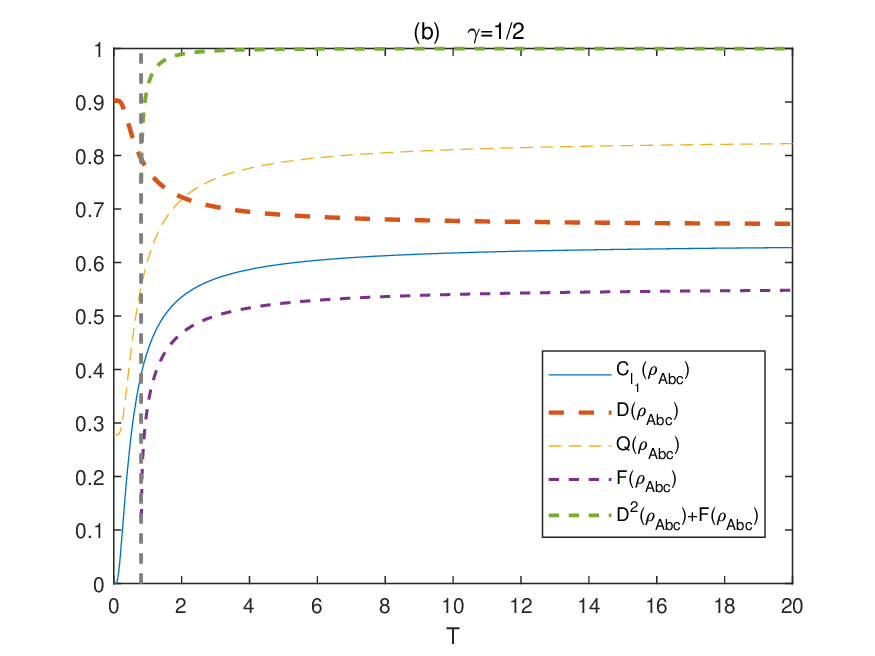}
    \end{minipage}
    \hfill
    \begin{minipage}[b]{0.4\textwidth}
        \includegraphics[width=\linewidth]{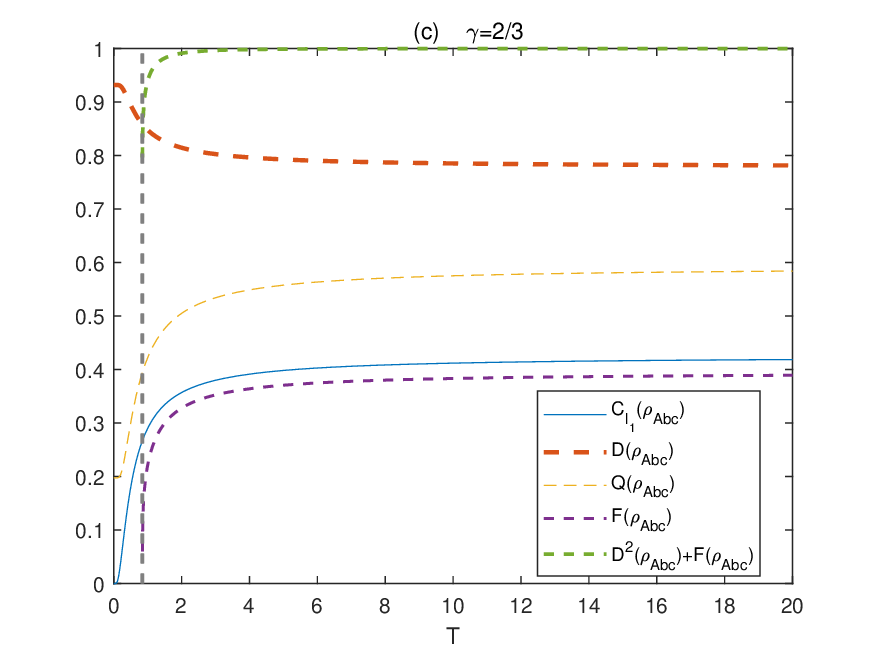}
    \end{minipage}
    \caption{The $l_{1}$-norm of quantum coherence $[C_{l_{1}}(\rho'_{Abc})]$, FOC $[D(\rho'_{Abc})]$, GC $[Q(\rho'_{Abc})]$, CF $[F(\rho'_{Abc})]$, and $D^{2}(\rho'_{Abc})+F(\rho'_{Abc})$ as functions of the Hawking temperature $T$ with $\omega=1$ for different $\gamma$: $\gamma=1/3$, $\gamma=1/2$ and $\gamma=2/3$.}
    \label{ADFigAbc}
\end{figure}

In Fig.\ref{ADFigAbc} (a-c), we depict the $C_{l_{1}}(\rho'_{Abc})$, $D(\rho'_{Abc})$, $Q(\rho'_{Abc})$, $F(\rho'_{Abc})$, and $D^{2}(\rho'_{Abc})+F(\rho'_{Abc})$ as functions of the Hawking temperature $T$ with $\omega=1$ for different $\gamma$. For $\gamma=\frac{1}{3}$, it is evident from Fig.\ref{ADFigAbc}(a) that $C_{l_{1}}(\rho'_{Abc})$, $Q(\rho'_{Abc})$ and $F(\rho'_{Abc})$ exhibit a monotonic increase as the Hawking temperature $T$ rises, whereas $D(\rho'_{Abc})$ demonstrates a monotonic decrease with the increase in Hawking temperature $T$. The function $F(\rho'_{Abc})$ is properly defined exclusively under the condition that $Q(\rho'_{Abc})\geq\frac{56}{81}$, and when $Q(\rho'_{Abc})=\frac{56}{81}$, $F(\rho'_{Abc})=0$. It is noted that the trade-off relationship between $D(\rho'_{Abc})$ and $F(\rho'_{Abc})$ gradually intensifies to reach a perfect trade-off as $T$ increases.
For $\gamma=\frac{1}{2}$ and $\gamma=\frac{2}{3}$, we see that the two corresponding subgraphs Fig.\ref{ADFigAbc} (b) and (c) share the same trend as Fig.\ref{ADFigAbc} (a).
For a fixed value of $T$, an increase in $\gamma$ leads to a decrease in $C_{l_{1}}(\rho'_{Abc})$, $Q(\rho'_{Abc})$ and $F(\rho'_{Abc})$, while causing $D(\rho'_{Abc})$ to increase.

Finally, from Eq.(\ref{rhoABc}) and  Eq.(\ref{ADchannel}) we derive
\begin{eqnarray}
\label{rho'ABC}
\rho'_{ABc}=\frac{1}{3}
\left (
\begin{array}{cccccccc}
c_{11}       &0                &0              &c_{14}            &0                &c_{16}                &0                       &0        \\
0            &c_{22}           &0              &0                 &0                &0                     &0                       &0         \\
0            &0                &c_{33}         &0                 &c_{35}           &0                     &0                       &c_{38}    \\
c_{41}       &0                &0              &c_{44}            &0                &c_{46}                &0                       &0         \\
0            &0                &c_{53}         &0                 &c_{55}           &0                     &0                       &0         \\
c_{61}       &0                &0              &c_{64}            &0                &c_{66}                &0                       &0         \\
0            &0                &0              &0                 &0                &0                     &c_{77}                  &0          \\
0            &0                &c_{83}         &0                 &0                &0                     &0                       &c_{88}
\end{array}
\right )
\end{eqnarray}
with
\begin{eqnarray*}
\begin{aligned}
&c_{11}=\gamma(\alpha^{2}+\beta^{2}\gamma)^{2}+\alpha^{2}+\gamma+\beta^{2}\gamma^{2},\\
&c_{22}=\gamma(1-\gamma)\beta^{2}(\alpha^{2}\gamma+\beta^{2}\gamma^{2}+1),\\
&c_{33}=\gamma(1-\gamma)\beta^{2}(\alpha^{2}\gamma+\beta^{2}\gamma^{2}+1)+1-\gamma,\\
&c_{44}=(1-\gamma)^{2}\beta^{2}(1+\beta^{2}\gamma),\\
&c_{55}=(1-\gamma)(\alpha^{2}+\beta^{2}\gamma)^{2},\\
&c_{66}=c_{77}=(1-\gamma)^{2}\beta^{2}(\alpha^{2}+\beta^{2}\gamma),\\
&c_{88}=(1-\gamma)^{3}\beta^{4},\\
\end{aligned}
\end{eqnarray*}
\begin{eqnarray}
\begin{aligned}
&c_{14}=c_{41}=(1-\gamma)\alpha\beta,\\
&c_{16}=c_{61}=(1-\gamma)\beta(\alpha^{2}+\beta^{2}\gamma),\\
&c_{35}=c_{53}=(1-\gamma)\alpha(\alpha^{2}+\beta^{2}\gamma),\\
&c_{38}=c_{83}=(1-\gamma)^{2}\beta^{3},\\
&c_{46}=c_{64}=(1-\gamma)^{2}\alpha\beta^{2}.
\end{aligned}
\end{eqnarray}

\begin{figure}[htbp]
    \centering
    \begin{minipage}[b]{0.4\textwidth} 
        \includegraphics[width=\linewidth]{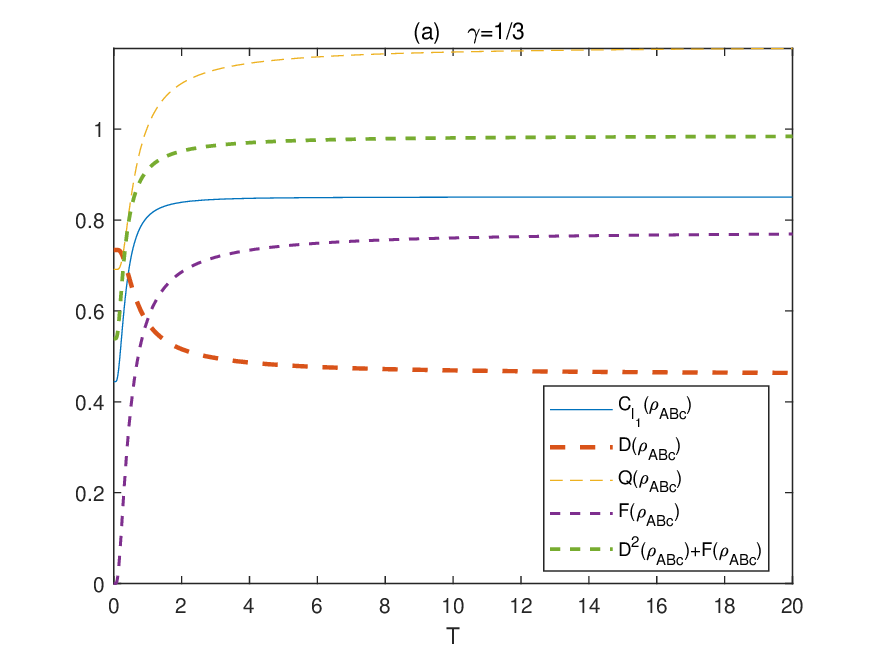} 

    \end{minipage}
    \hfill
    \begin{minipage}[b]{0.4\textwidth}
        \includegraphics[width=\linewidth]{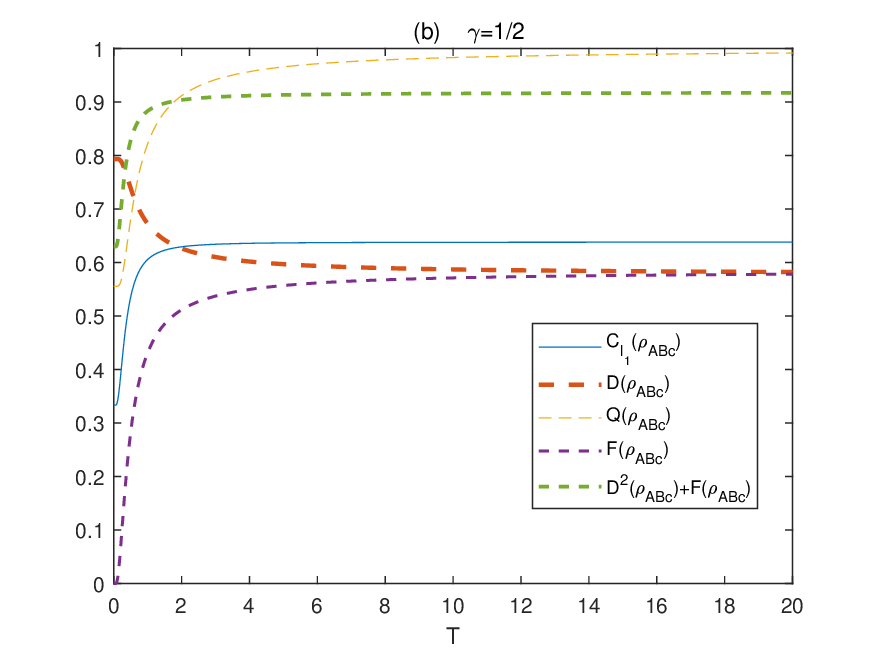}
    \end{minipage}
    \hfill
    \begin{minipage}[b]{0.4\textwidth}
        \includegraphics[width=\linewidth]{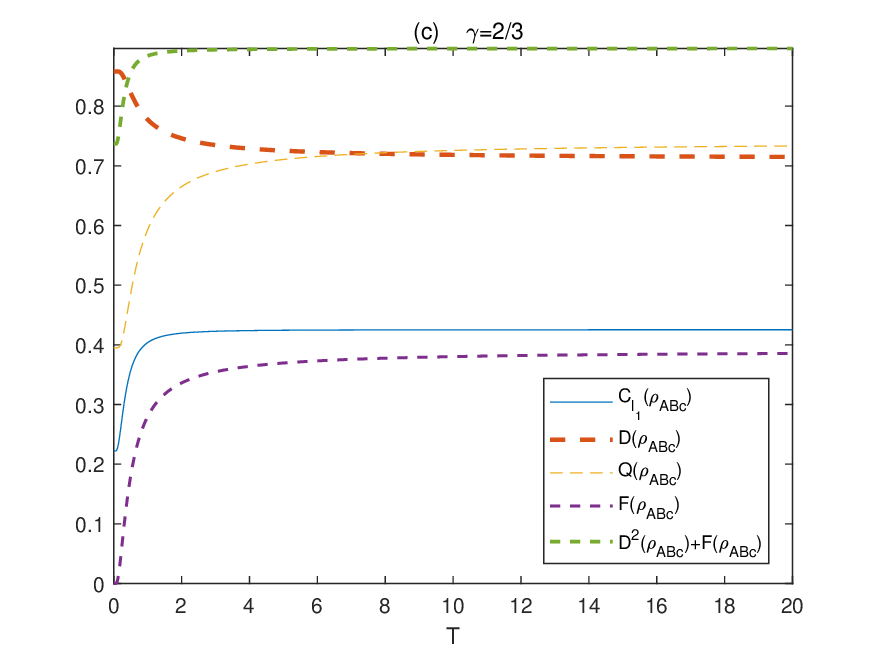}
    \end{minipage}
    \caption{The $l_{1}$-norm of quantum coherence $[C_{l_{1}}(\rho'_{ABc})]$, FOC $[D(\rho'_{ABc})]$, GC $[Q(\rho'_{ABc})]$, CF $[F(\rho'_{ABc})]$, and $D^{2}(\rho'_{ABc})+F(\rho'_{ABc})$ as functions of the Hawking temperature $T$ with $\omega=1$ for different $\gamma$: $\gamma=1/3$, $\gamma=1/2$ and $\gamma=2/3$.}
    \label{ADFigABc}
\end{figure}

In Fig.\ref{ADFigABc}(a-c), we show $C_{l_{1}}(\rho'_{ABc})$, $D(\rho'_{ABc})$, $Q(\rho'_{ABc})$, $F(\rho'_{ABc})$, and $D^{2}(\rho'_{ABc})+F(\rho'_{ABc})$ as functions of the Hawking temperature $T$ with $\omega=1$ for different $\gamma$. From Fig.\ref{ADFigABc}(a-c), it is seen that all the subgraphs exhibit the same trend: $C_{l_{1}}(\rho'_{ABc})$, $Q(\rho'_{ABc})$ and $F(\rho'_{ABc})$ monotonically increase as $T$ increases, while $D(\rho'_{ABc})$ increases as $T$ increases.
Notably, $D^{2}(\rho'_{ABc})+F(\rho'_{ABc})$ increases monotonically from its minimum value, asymptotically approaching a stable value, without ever reaching a perfect trade-off relationship between $D(\rho'_{ABc})$ and $F(\rho'_{ABc})$, i.e., $D^{2}(\rho'_{ABc})+F(\rho'_{ABc})<1$. For a given $T$, increasing $\gamma$ reduces $C_{l_{1}}(\rho'_{ABc})$, $Q(\rho'_{ABc})$ and $F(\rho'_{ABc})$, but increases $D(\rho'_{ABc})$.

\section{Conclusion}
We have investigated certain quantumness in the vicinity of the Schwarzschild black hole by utilizing the W state. On the one hand, we have studied the influence of the Hawking effect on $l_{1}$-norm of quantum coherence, FOC, CF and GC in Schwarzschild black hole, by assuming that Alice is located in an asymptotically flat region, while Bob and Charlie hover near the event horizon of the Schwarzschild black hole. Then, we have explored the following three scenarios: systems with three physically accessible modes, systems with two physically accessible modes and systems with one physically accessible mode. For the first scenario, in~\cite{Ali.2024}, the authors showed that Hawking temperatures degrade entanglement based on the GHZ states, however, we find that the Hawking effect of the black hole not only disrupts but also enhance the quantum entanglement based on the W states, while destroying the quantum coherence. For the other two scenarios, the Hawking temperature $T$ exerts a positive influence on quantum coherence and quantum entanglement. On the other hand, we have explored the influence of both the Hawking effect and environmental noise (AD channels) on $l_{1}$-norm of quantum coherence, FOC, CF and GC. In a completely physically accessible scenario, we conclude that the Hawking effect of the black hole disrupts quantum coherence but exerts a positive influence on quantum entanglement under the AD channels. Our results may highlight further investigations on other quantumness near the Schwarzschild black hole based on other quantum states.

\begin{acknowledgments}
This work is supported by the National Natural Science Foundation of China (NSFC) under Grant No. 12171044; the Natural Science Foundation of Hainan Province under Grant No. 125RC744 and the Hainan Academician Workstation (J\"urgen Jost).
\end{acknowledgments}

\appendix
\section{APPENDIX A: The FOC, CF and GC of states $\rho_{ABC}$, $\rho_{Abc}$ and $\rho_{ABc}$}
\label{App1}
The states with respect to all subsystems are, respectively,
\begin{eqnarray}
\rho_{A}=\frac{1}{3}
\left (
\begin{array}{cc}
2                 &0                \\
0                 &1
\end{array}
\right ),
\end{eqnarray}
\begin{eqnarray}
\rho_{B}=\frac{1}{3}
\left (
\begin{array}{cc}
2\alpha^{2}       &0                \\
0                 &1+2\beta^{2}
\end{array}
\right ),
\end{eqnarray}
\begin{eqnarray}
\rho_{b}=\frac{1}{3}
\left (
\begin{array}{cc}
1+2\alpha^{2}     &0               \\
0                 &2\beta^{2}
\end{array}
\right ),
\end{eqnarray}
\begin{eqnarray}
\rho_{C}=\frac{1}{3}
\left (
\begin{array}{cc}
2\alpha^{2}       &0                \\
0                 &1+2\beta^{2}
\end{array}
\right ),
\end{eqnarray}
and
\begin{eqnarray}
\rho_{c}=\frac{1}{3}
\left (
\begin{array}{cc}
1+2\alpha^{2}     &0               \\
0                 &2\beta^{2}
\end{array}
\right ).
\end{eqnarray}
Obviously, $\rho_{B}=\rho_{C}$ and $\rho_{b}=\rho_{c}$.

Then, in accordance with $D(\rho_{i})=\sqrt{2tr(\rho_{i}^{2})-1}$ the FOC of individual reduced states are
\begin{eqnarray}
\begin{aligned}
D(\rho_{A})=\sqrt{2\times(\frac{4}{9}+\frac{1}{9})-1}=\frac{1}{3},
\end{aligned}
\end{eqnarray}
\begin{eqnarray}
\begin{aligned}
D(\rho_{B})&=\sqrt{2\times\left[\frac{4}{9}\alpha^{2}+\frac{1}{9}(1+2\beta^{2})^{2}\right]-1}\\
&=\sqrt{\frac{8}{9}(\alpha^{4}+\beta^{2}+\beta^{4})-\frac{7}{9}}\\
&=\frac{1}{3}\sqrt{8(\alpha^{4}+\beta^{2}+\beta^{4})-7},
\end{aligned}
\end{eqnarray}
\begin{eqnarray}
\begin{aligned}
D(\rho_{b})&=\sqrt{2\times\left[\frac{1}{9}(1+2\alpha^{2})^{2}+\frac{4}{9}\beta^{2}\right]-1}\\
&=\sqrt{\frac{8}{9}(\alpha^{4}+\alpha^{2}+\beta^{4})-\frac{7}{9}}\\
&=\frac{1}{3}\sqrt{8(\alpha^{4}+\alpha^{2}+\beta^{4})-7}.
\end{aligned}
\end{eqnarray}
Notice that $\rho_{B}=\rho_{C}$ and $\rho_{b}=\rho_{c}$, then $D(\rho_{B})=D(\rho_{C})$ and $D(\rho_{b})=D(\rho_{c})$.

According to Eq.(\ref{FOC}), we have
\begin{eqnarray*}
\begin{aligned}
D(\rho_{ABC})&=\sqrt{\frac{1}{3}\times\left[\frac{1}{9}+\frac{16}{9}(\alpha^{4}+\beta^{2}+\beta^{4})-\frac{14}{9}\right]}\\
&=\frac{1}{3}\sqrt{\frac{16(\alpha^{4}+\beta^{2}+\beta^{4})-13}{3}},
\end{aligned}
\end{eqnarray*}
\begin{eqnarray*}
\begin{aligned}
D(\rho_{Abc})&=\sqrt{\frac{1}{3}\times\left[\frac{1}{9}+\frac{16}{9}(\alpha^{4}+\alpha^{2}+\beta^{4})-\frac{14}{9}\right]}\\
&=\frac{1}{3}\sqrt{\frac{16(\alpha^{4}+\alpha^{2}+\beta^{4})-13}{3}},
\end{aligned}
\end{eqnarray*}
and
\begin{eqnarray*}
\begin{aligned}
D(\rho_{ABc})&=\sqrt{\frac{1}{3}\times\left[\frac{1}{9}+\frac{8}{9}(2\alpha^{4}+2\beta^{4}+1)-\frac{14}{9}\right]}\\
&=\frac{1}{3}\sqrt{\frac{16(\alpha^{4}+\beta^{4})-5}{3}}.
\end{aligned}
\end{eqnarray*}

Subsequently, from $C_{i(jk)}=2\sqrt{det(\rho_{i})}$, $\rho_{B}=\rho_{C}$ and $\rho_{b}=\rho_{c}$, we obtain
\begin{eqnarray*}
\begin{aligned}
C_{A(BC)}=C_{A(bc)}=C_{A(Bc)}=\frac{2\sqrt{2}}{3},
\end{aligned}
\end{eqnarray*}
\begin{eqnarray*}
\begin{aligned}
C_{B(CA)}=C_{C(AB)}=C_{B(cA)}=\frac{2}{3}\sqrt{2\alpha^{2}(1+2\beta^{2})},
\end{aligned}
\end{eqnarray*}
and
\begin{eqnarray*}
\begin{aligned}
C_{b(cA)}=C_{c(Ab)}=C_{c(AB)}=\frac{2}{3}\sqrt{2\beta^{2}(1+2\alpha^{2})}.
\end{aligned}
\end{eqnarray*}
Therefore, by Eq.(\ref{GC}) we derive
\begin{eqnarray*}
\begin{aligned}
Q(\rho_{ABC})&=\frac{1}{2}\times\left\{\frac{8}{9}+2\times\frac{4}{9}\left[2\alpha^{2}(1+2\beta^{2})\right]\right\}\\
&=\frac{4}{9}\left[1+2\alpha^{2}(1+2\beta^{2})\right],
\end{aligned}
\end{eqnarray*}
\begin{eqnarray*}
\begin{aligned}
Q(\rho_{Abc})&=\frac{1}{2}\times\left\{\frac{8}{9}+2\times\frac{4}{9}\left[2\beta^{2}(1+2\alpha^{2})\right]\right\}\\
&=\frac{4}{9}\left[1+2\beta^{2}(1+2\alpha^{2})\right]
\end{aligned}
\end{eqnarray*}
and
\begin{eqnarray*}
\begin{aligned}
Q(\rho_{ABc})&=\frac{1}{2}\times\left[\frac{8}{9}+\frac{4}{9}\left(2\alpha^{2}+2\beta^{2}+8\alpha^{2}\beta^{2}\right)\right]\\
&=\frac{8}{9}\left(1+2\alpha^{2}\beta^{2}\right).
\end{aligned}
\end{eqnarray*}

Furthermore, from Eq.(\ref{CF}) we get
\begin{eqnarray*}
\begin{aligned}
F(\rho_{ABC})&=\left\{\frac{16}{3}Q(\rho_{ABC})\times\left[Q(\rho_{ABC})-\frac{8}{9}\right]\times\frac{4}{9}\times\frac{4}{9}\right\}^{\frac{1}{4}}\\
&=\frac{4}{3}\left\{\frac{1}{3}Q(\rho_{ABC})\times\left[Q(\rho_{ABC})-\frac{8}{9}\right]\right\}^{\frac{1}{4}},
\end{aligned}
\end{eqnarray*}
\begin{eqnarray*}
\begin{aligned}
F(\rho_{Abc})&=\left\{\frac{16}{3}Q(\rho_{Abc})\times\left[Q(\rho_{Abc})-\frac{8}{9}\right]\times\frac{4}{9}\times\frac{4}{9}\right\}^{\frac{1}{4}}\\
&=\frac{4}{3}\left\{\frac{1}{3}Q(\rho_{Abc})\times\left[Q(\rho_{Abc})
-\frac{8}{9}\right]\right\}^{\frac{1}{4}}
\end{aligned}
\end{eqnarray*}
and
\begin{eqnarray*}
\begin{aligned}
F(\rho_{ABc})&=\left\{\frac{16}{3}Q(\rho_{ABc})\times\left[Q(\rho_{ABc})-\frac{8}{9}\right]\right.\\
&\left.\times\left(\frac{8}{9}-\frac{8}{9}\alpha^{2}\right)\times\left(\frac{8}{9}-\frac{8}{9}\beta^{2}\right)\right\}^{\frac{1}{4}}\\
&=\left(\frac{16}{3}Q(\rho_{ABc})\times\frac{16}{9}\alpha^{2}\beta^{2}\times\frac{8}{9}\alpha^{2}\times\frac{8}{9}\beta^{2}\right)^{\frac{1}{4}}\\
&=\frac{4}{3}\alpha\beta\left[\frac{64}{27}Q(\rho_{ABc})\right]^{\frac{1}{4}}.
\end{aligned}
\end{eqnarray*}

\section{APPENDIX B: The states of the subsystems under the AD channels}

The states of all subsystems evolving under the AD channels can be expressed as
\begin{eqnarray}
\rho'_{A}=\frac{1}{3}
\left (
\begin{array}{cc}
2+\gamma          &0                \\
0                 &1-\gamma
\end{array}
\right ),
\end{eqnarray}
\begin{eqnarray}
\rho'_{B}=\frac{1}{3}
\left (
\begin{array}{cc}
2\alpha^{2}+\gamma(1+2\beta^{2})       &0                           \\
0                                      &(1-\gamma)(1+2\beta^{2})
\end{array}
\right ),
\end{eqnarray}
\begin{eqnarray}
\rho'_{b}=\frac{1}{3}
\left (
\begin{array}{cc}
1+2\alpha^{2}+2\beta^{2}\gamma     &0                        \\
0                                  &2\beta^{2}(1-\gamma)
\end{array}
\right ),
\end{eqnarray}
\begin{eqnarray}
\rho'_{C}=\frac{1}{3}
\left (
\begin{array}{cc}
2\alpha^{2}+\gamma(1+2\beta^{2})       &0                           \\
0                                      &(1-\gamma)(1+2\beta^{2})
\end{array}
\right )
\end{eqnarray}
and
\begin{eqnarray}
\rho'_{c}=\frac{1}{3}
\left (
\begin{array}{cc}
1+2\alpha^{2}+2\beta^{2}\gamma     &0                        \\
0                                  &2\beta^{2}(1-\gamma)
\end{array}
\right ).
\end{eqnarray}
Obviously, $\rho'_{B}=\rho'_{C}$ and $\rho'_{b}=\rho'_{c}$.


\end{document}